\documentclass[10pt,aps,superscriptaddress,prb]{revtex4}
\usepackage{epsfig}
\usepackage{amsmath}
\usepackage{amssymb}
\usepackage{bbold}

\usepackage{xcolor}
\usepackage{verbatim}  
\usepackage{mathrsfs} 
\usepackage{marvosym} 
\usepackage{leftidx}  
\usepackage{pifont}  

\usepackage[colorlinks]{hyperref}
\newcommand\be{\begin{equation}}
\newcommand\ee{\end{equation}}
\newcommand\bea{\begin{eqnarray}}
\newcommand\eea{\end{eqnarray}}

\newcommand{\fatalpha}{{\bf \alpha \kern -0.44em \alpha}}
\newcommand{\fatsigma}{{\bf \sigma \kern -0.54em \sigma}}
\newcommand{\tpchi}{{\bf D \kern -0.35em D}}
\newcommand{\llambda}{{\bf \lambda \kern -0.45em \lambda}}

\bibliography{plain}

\begin{document}
\title{Bound states for massive Dirac fermions in graphene in a magnetic step field}
\author{L. Dell'Anna}
\affiliation{Dipartimento di Fisica e Astronomia, University of Padova, via F. Marzolo 8, 35151 Padova, Italy}
\author{A. Alidoust Ghatar}
\author{D. Jahani}
\affiliation{Material and Energy Research Center, Tehran, Iran}



\begin{abstract}

We calculate the spectrum of massive Dirac fermions in graphene in the presence of an inhomogeneous magnetic field modeled by a step function. We find an analytical universal relation between the bandwidths and the propagating velocities of the modes at the border of the magnetic region, showing how by tuning the mass term one can control the speed of these traveling edge states. 

\end{abstract}

\maketitle

\section{ Introduction}


Graphene is generally described by massless Dirac fermions \cite{h1}, nevertheless different techniques have been developed for nanotechnological applications and for exploring non-trivial topological properties, in order to generate a gap at the Dirac points \cite{m1,m2,m3} so to include a mass term in the Dirac-Weyl Hamiltonian, which describes the low energy physics in graphene \cite{h1,h2,h3,h4}. Mass terms, confining scalar potentials or magnetic fields can spoil the simple linear dispersion of the original massless Dirac fermions. \cite{h5,h6,h7,h8,h9,h10,h11,h12}

In particular, applying inhomogeneous magnetic fields perpendicularly to the graphene sheet one can produce bound states trapped in the vicinity of the discontinuity of the magnetic field and propagating along the magnetic edges. \cite{h14,h14b,h13,h15}
Several bound state spectra have been already obtained and scattering problems solved for massless Dirac fermions in graphene embedded in discontinuous magnetic fields employing boundary conditions. \cite{h13,h14,h14b,h15,h16,h17,h18,h19,h20,h21,h22,h23,h24,h25,h26,h27} 

What we are going to present in this work, instead, is the bound state spectrum for massive Dirac fermions in the simplest inhomogeneous magnetic pattern described by a step function. In the massless limit we recover the known results. \cite{h13,h14} 
We observe some universal behaviors, in particular we show analytically that, in this magnetic structure, for each band, the bound state threshold does not depend on the mass term even if the energy levels do depend on it. Moreover we show that the maximum of the velocity of the edge modes propagating along the magnetic boundary, as a function of the mass term, seems to be proportional to the bandwidth, therefore the ratio between the bandwidth and the corresponding maximum velocity is mass-independent quantity. In summary, by this analysis, we show how, by tuning the mass term, one can control the propagating speed of the modes located at the edge of a magnetic region, relevant for future tunable graphene-based mesoscopic devices.

\section{Magnetic step}\label{section.two}

Let us consider a magnetic field perpendicular to the plane of graphene, $z$-direction, and with a step profile along one direction in the plane of graphene, $B_z(x)=B\,\theta(x)$, where $\theta(x)$ is the Heaviside theta function. 
 The potential vector is, therefore, $\vec{A}=(0,A(x),0)$, defined by 
   \begin{equation}\label{eq1}
  A(x)=\frac{\hbar c}{el_{B}^{2}}x\,\theta(x)
\end{equation}
 where $\hbar=h/2\pi$ with $h$ the Planck constant, $c$ the speed of light, $e$ the elementary charge, and $l_{B}=\sqrt{\hbar c/eB}$ the magnetic length.
 We are supposing that the length scale over which $B(x)$ significantly varies, say $\lambda_B$, is assumed much larger than the lattice spacing so that, at low energy scales, the two Dirac points in the massless limit are not coupled by the magnetic field and can be treated separately. We assume also that $\lambda_B$ is much smaller than the quasiparticle Fermi wavelength so that we can safely approximate $B(x)$ as a step function.
The Dirac-Weyl equation is, then given by 
 \begin{equation}
 \label{DW}
 \Big(\upsilon_{F}(\sigma_{x}\pi_{x}+\sigma_{y}\pi_{y})+\Delta\sigma_{z}\Big)\psi(x,y)=\varepsilon\psi(x,y)
 \end{equation}
with $\upsilon_{F}$ the Fermi velocity, $\sigma_{x} , \sigma_{y},  \sigma_{z}$ Pauli matrices and  $\pi$ the
momentum operator 
 $\vec \pi=-i\hbar \vec \nabla+\frac{e}{c}\vec A$.
 Because of the translational invariance along the $y$-direction, the spinor can be written as
 \begin{equation}
 	\psi(x,y)=e^{i k_{y}y}\,\psi(x)\equiv e^{i k_{y}y}\,\left(
 	\begin{array}{c}
 		a(x) \\
 		b(x) \\
 	\end{array}
 	\right)
 \end{equation}
 Let us split the space in two regions, I and II.\\
\paragraph*{Region I.}For $x>0$, from Eq.~(\ref{DW}) we can write the following equations for the two components 
   \begin{eqnarray}
   \label{eq3}
&& \Delta\, a(x)-i\upsilon_{F}\left[\hbar\frac{d}{dx}+\left(\hbar k_{y}+\frac{eA(x)}{c}\right)\right]b(x)=\varepsilon \,a(x)\\
&&-i\upsilon_{F}\left[\hbar\frac{d}{dx}-\left(\hbar k_{y}+\frac{eA(x)}{c}\right)\right]a(x)-\Delta\, b(x)=\varepsilon \,b(x)
\end{eqnarray}
From these equations, putting one into the other, we obtain
 \begin{equation}
\left[l_{B}^{2}\hbar^{2}\frac{d^{2}}{dx^{2}}-\left(\hbar k_{y}l_{B}+\left(\frac{el_{B}}{c}\right)A(x)\right)^{2}-
\left(\frac{l_{B}^{2}e}{c}\right)\left(\frac{dA(x)}{dx}\right)+\frac{l_{B}^{2}}{\upsilon_{F}^{2}}\left(\varepsilon^{2}-\Delta^{2}\right)\right]a(x)=0
\end{equation}
Using Eq.~\eqref{eq1}, we can write the following equation, for $x>0$,
\begin{equation}\label{eq6}
  \left[\frac{d^{2}}{d\left(\frac{x}{l_{B}}\right)^{2}}-\left(k_{y}l_{B}+x/l_{B}\right)^{2}-1
  +\frac{l_{B}^{2}}{\hbar^{2}\upsilon_{F}^{2}}
  \left(\varepsilon^{2}-\Delta^{2}\right)\right]a(x)=0
\end{equation}
Notice that this is an effective one-dimensional Schr\"odinger equation where for $k_y<0$ the potential develops a minimum within the magnetic region for which there exist bound state solutions with energy $|k_y|>\sqrt{\left(\varepsilon^{2}-\Delta^{2}\right)}/\hbar \upsilon_F$, as we will verify in what follows. Making the change of variables 
\be 
\xi=k_{y}l_{B}+x/l_{B},
\ee  
Eq.~\eqref{eq6} becomes simply
\begin{equation}
\label{eqbx}
	\left(\frac{d^{2}}{d\xi^2}-\xi^{2}-1+2\eta \right)a(x)=0
\end{equation}
where we defined the quantity, which is generally a real number,
\begin{equation}
\label{enne}
  \eta =\frac{l_{B}^{2}}{2\hbar^{2}\upsilon_{F}^{2}}\left(\varepsilon^{2}-\Delta^{2}\right)
\end{equation}
The normalizable solution of Eq.~\eqref{eqbx}  is 
\be
a(x)= c_{I}\,D_{\eta -1}(\sqrt{2}\,\xi)
\ee
where $D_{\eta}(z)$ is a parabolic cylinder function and $c_I$ a constant value. Notice that if $\eta=n$ is a non-negative integer number, one can write $D_{n}(\sqrt{2}\xi)=2^{-n/2}e^{-\xi^{2}/2} H_{n}(\xi)$, namely in terms of the Hermite polynomials
  $H_{n}(\xi)=(-1)^{n} e^{\xi^{2}} \frac{d^{n}}{d\xi^{n}} e^{-\xi^{2}}$. 
To find the second component of the spinor we can write 
\be
(\varepsilon +\Delta) b(x)=-i\frac{\hbar \upsilon_F}{l_B}\left(\frac{d}{d\xi}-\xi\right) a(x)
\ee
and using the recursive relation
\be
\frac{d}{dz}D_{\eta-1}(z)-\frac{z}{2}D_{\eta-1}(z)+D_{\eta}(z)=0
\ee
we get the complete spinorial wavefunction
  \begin{equation}
  \label{phi1}
    \psi(x)=c_I\,\left(
                                    \begin{array}{c}
                                      D_{\eta-1}(\sqrt{2}\, \xi) \\
                                      \frac{i \sqrt{2}\,\hbar \upsilon_F}{(\varepsilon+\Delta)l_B}D_{\eta}(\sqrt{2}\,\xi) \\
                                    \end{array}
                                  \right). 
 \end{equation}

\paragraph*{Region II.}
 For $x<0$, we obtain the following equation by placing $A(x)=0$ in Eq.~\eqref{eq3}
  \begin{eqnarray}
  \label{caseII1}
&& \frac{\left(\varepsilon-\Delta\right)}{\hbar\upsilon_{F}}a(x)+i\left(\frac{d}{dx}+ k_{y}\right)b(x)=0\\
&& \frac{\left(\varepsilon+\Delta\right)}{\hbar\upsilon_{F}}b(x)+i\left(\frac{d}{dx}- k_{y}\right)a(x)=0
\label{caseII2}
\end{eqnarray}
so that, analogously to what done in the other case, we can write
\begin{equation}
  \left[\frac{d^{2}}{dx^2}-k_y^2
  +\frac{\left(\varepsilon^{2}-\Delta^{2}\right)}{\hbar^{2}\upsilon_{F}^{2}}
  \right]a(x)=0
\end{equation}
whose solution is can be written as
\begin{equation}
\label{ax}
  a(x)=c_{II} \left(e^{k_{x} x}+r\, e^{-k_{x} x}\right)
\end{equation}
with $r$ a constant value and where we defined 
\be
\label{kx}
k_x=\sqrt{k_y^2-\left(\varepsilon^2-\Delta^2\right)/\hbar^2\upsilon_F^2}.
\ee
The bound state solutions are those with $k_y^2>\left(\varepsilon^2-\Delta^2\right)/\hbar^2\upsilon_F^2$ so that, for $x<0$ the only normalizable contribution in Eq.~(\ref{ax}) is the first one, $a(x)=c_{II} e^{k_{x}x}$. 
Using Eq.~(\ref{caseII2}) we get the other component of the spinor, $b(x)=-i\frac{\hbar\upsilon_{F}}{(\varepsilon+\Delta)}\left(\frac{d}{dx}- k_{y}\right)a(x)$, getting the following bound-state wavefunction
\begin{equation}
\label{phi2}
\psi(x)=c_{II}\,e^{k_x x}\left(
                                    \begin{array}{c}
                                     1 \\
                                      \frac{i\hbar\upsilon_{F}(k_y-k_x)}{(\varepsilon+\Delta)} \\
                                    \end{array}
                                  \right). 
 \end{equation}
Imposing the matching condition at $x=0$, from Eqs.~(\ref{phi1}) and (\ref{phi2}) we find
\begin{equation}
\label{eq27}
 l_{B}\left(k_y-k_x\right) D_{(\varepsilon^{2}l_{B}^{2}-\Delta^{2}l_{B}^{2})/(2\hbar^{2}\upsilon_{F}^{2})-1}(\sqrt{2}\,k_{y}l_{B})
 = \sqrt{2}\,D_{(\varepsilon^{2}l_{B}^{2}-\Delta^{2}l_{B}^{2})/(2\hbar^{2}\upsilon_{F}^{2})}~(\sqrt{2}\,k_{y}l_{B})
\end{equation}
where we used Eq.~(\ref{enne}) and where $k_x$ is defined in Eq.~(\ref{kx}).\\
If we put $\Delta = 0$ in Eq.~\eqref{eq27}, the matching condition reduces to that of the gapless graphene \cite{h14}. 
In this case in addition to the finite-energy states, solution of the above equation, there is also the zero energy state, $\tilde \epsilon_0=0$, for $k_y<0$, whose wavefunction is 
\be
\label{zeromode}
\psi_0(x)=\left(\begin{array}{c}0\\1 \end{array}\right)\left(\theta(-x)+\theta(x)e^{-x^2/2}\right)e^{-k_y x}.
\ee
In order to find the finite-energy spectrum defined by Eq.~(\ref{eq27}), in the general case of gapped graphene, it is convenient to introduce the dimensionless parameters 
\begin{eqnarray}
\label{eq28}
&&\tilde \epsilon^2=(\varepsilon^{2}l_{B}^{2})/(\upsilon_{F}^{2}\hbar^{2})\\
&&\tilde \Delta^2=(\Delta^{2}l_{B}^{2})/(\upsilon_{F}^{2}\hbar^{2})\\
&&\tilde k_y=k_{y}l_{B}
\end{eqnarray}
such that Eq.~\eqref{eq27} can be written as follows 
\begin{equation}
\label{mach_cond}
\left(\tilde k_y-\sqrt{{\tilde k_y}^{2}- (\tilde\epsilon^2-\tilde \Delta^2)}\right)  D_{(\tilde \epsilon^2-\tilde \Delta^2)/2-1}(\sqrt{2}\,\tilde k_{y})=\sqrt{2}\,D_{(\tilde\epsilon^2-\tilde\Delta^2)/2}(\sqrt{2}\,\tilde k_{y})
\end{equation}
whose solutions are quantized, $\tilde \epsilon_n$, with $n=1,2,3,...$. 
Notice that Eq.~(\ref{mach_cond}) is valid for $\tilde \epsilon\neq -\tilde \Delta$. 
Also in this case there is an extra-state with a completely flat band at $\tilde \epsilon_0=- \tilde \Delta$ whose wavefunctions is described by  Eq.~(\ref{zeromode}), localized at the edge, where the discontinuity of the magnetic field is located, but whose band is not dispersive. 

 \section{Results}
  \label{section.three}
\begin{figure}[h!]
\centering
\includegraphics[width=0.25\textwidth]{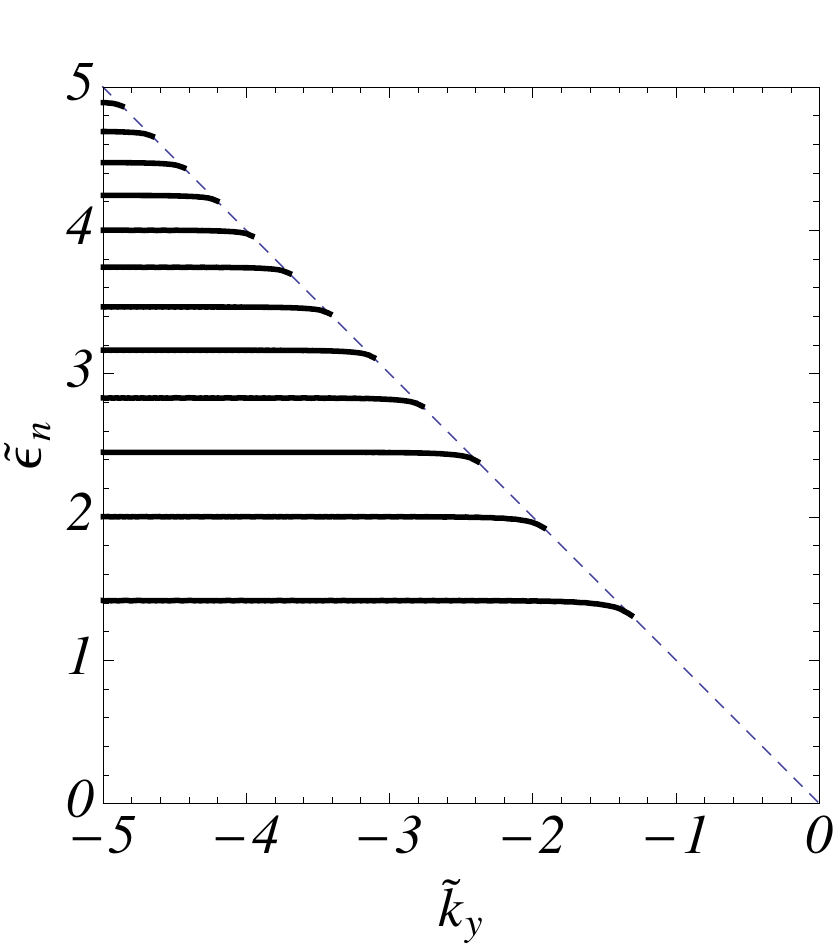}\,
\includegraphics[width=0.25\textwidth]{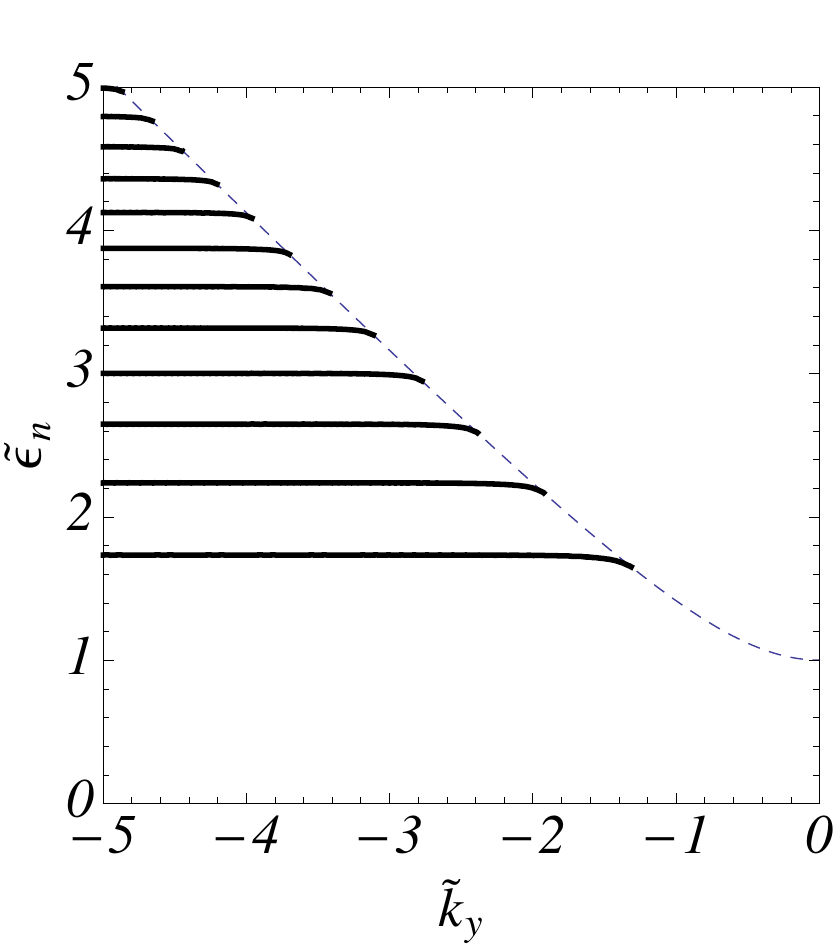}\,
\includegraphics[width=0.25\textwidth]{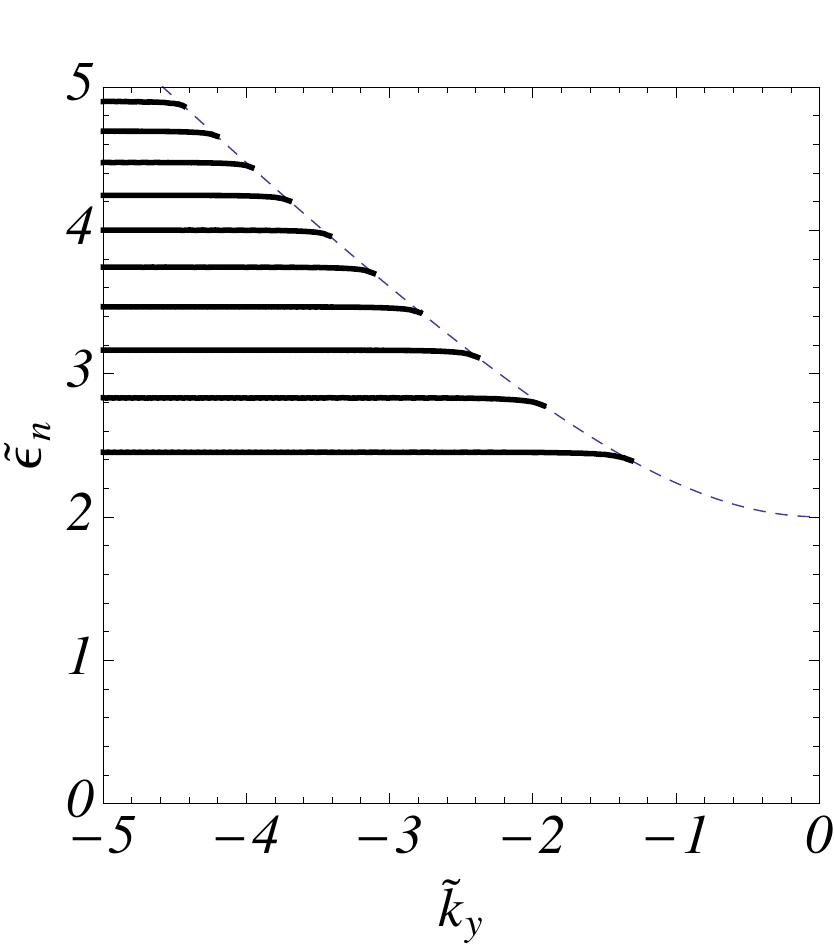}
\caption{Spectrum of the low-lying positive-energy edge states at a magnetic step for $\tilde \Delta=0$ (left), $\tilde \Delta=1$ (middle) and $\tilde \Delta=2$ (right). The dashed line denotes the threshold value for edge state solutions, $\tilde \epsilon=\sqrt{\tilde k_y^2+\tilde \Delta^2}$. For large negative $\tilde k_y$ the levels rapidly approach the Landau level values $\sqrt{{\tilde\Delta}^2+2n}$, with $n\ge1$ positive integer numbers. 
The negative-energy spectrum is specular with, in addition, the flat zeroth energy level $\tilde\epsilon_0=-\tilde \Delta$. }
\label{fig.spectra}
\end{figure}

\begin{figure}[h!]
\centering
\includegraphics[width=0.25\textwidth]{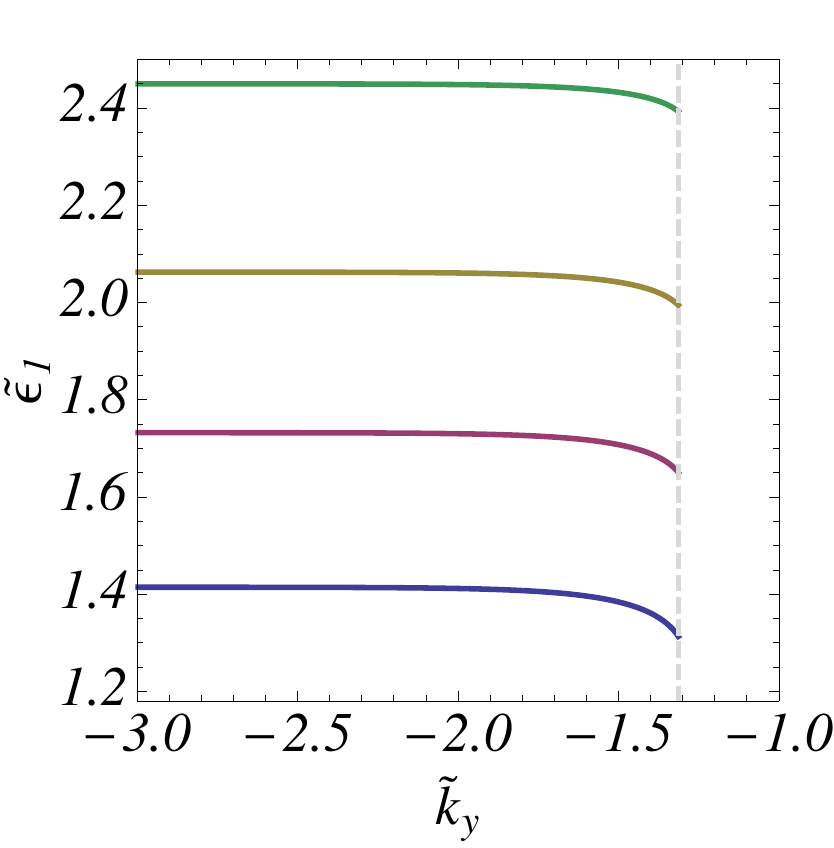}\,
\includegraphics[width=0.25\textwidth]{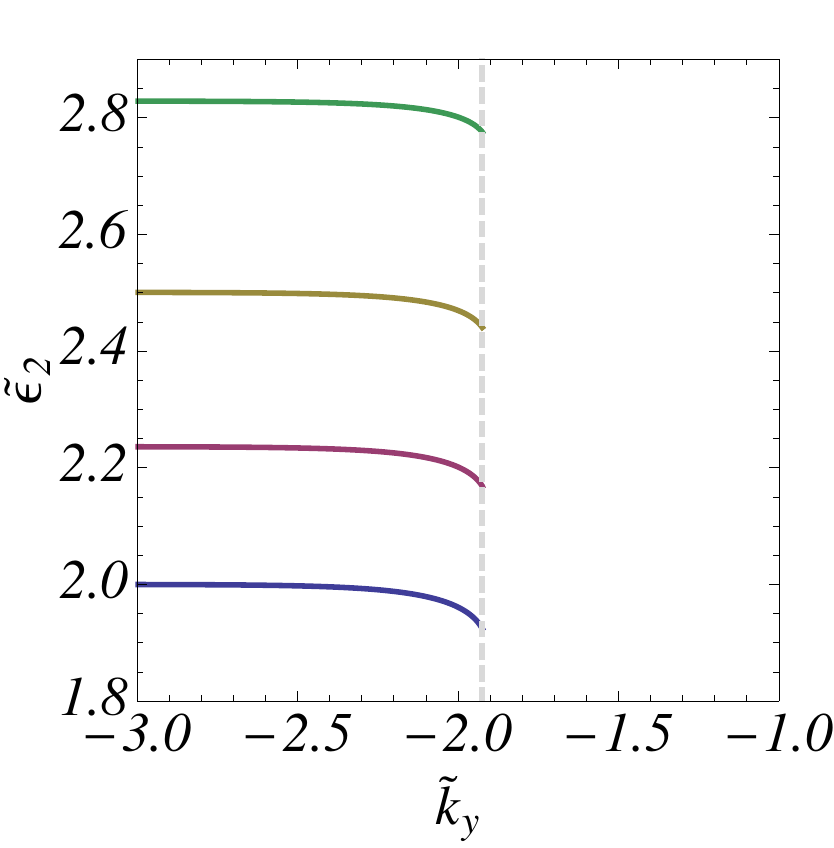}\,
\includegraphics[width=0.25\textwidth]{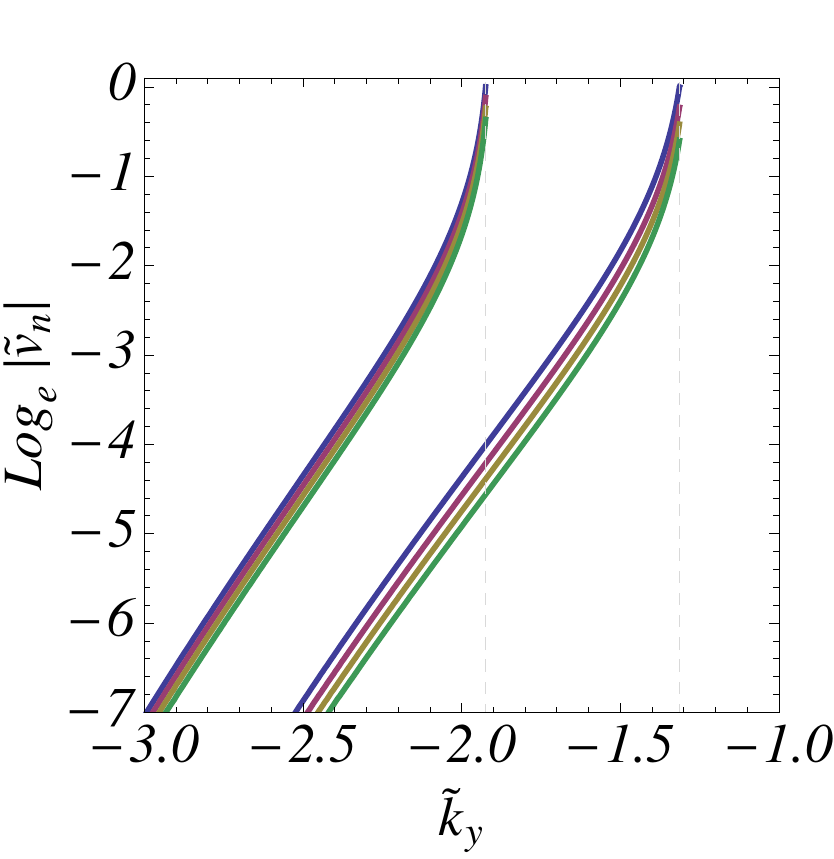}
\caption{First energy level $\tilde \epsilon_1$ (left) and second energy level $\tilde \epsilon_2$ (middle), solutions of Eq.~(\ref{mach_cond}), for increasing values of the mass term, $\tilde \Delta=0$ (blue line), $\tilde \Delta=1$ (red line), $\tilde \Delta=1.5$ (yellow line), $\tilde \Delta=2$ (green line). (Right) Modulus of the velocities, in log-scale, associated to the first and the second energy levels, defined as $\tilde v_n=\partial_{\tilde k_y}\tilde \epsilon_n$, for the same values of $\tilde \Delta$ as in the first two plots, $\tilde \Delta=0, 1, 1.5, 2$.}
\label{fig.en}
\end{figure}

\begin{figure}[h!]
\centering
\includegraphics[width=0.25\textwidth]{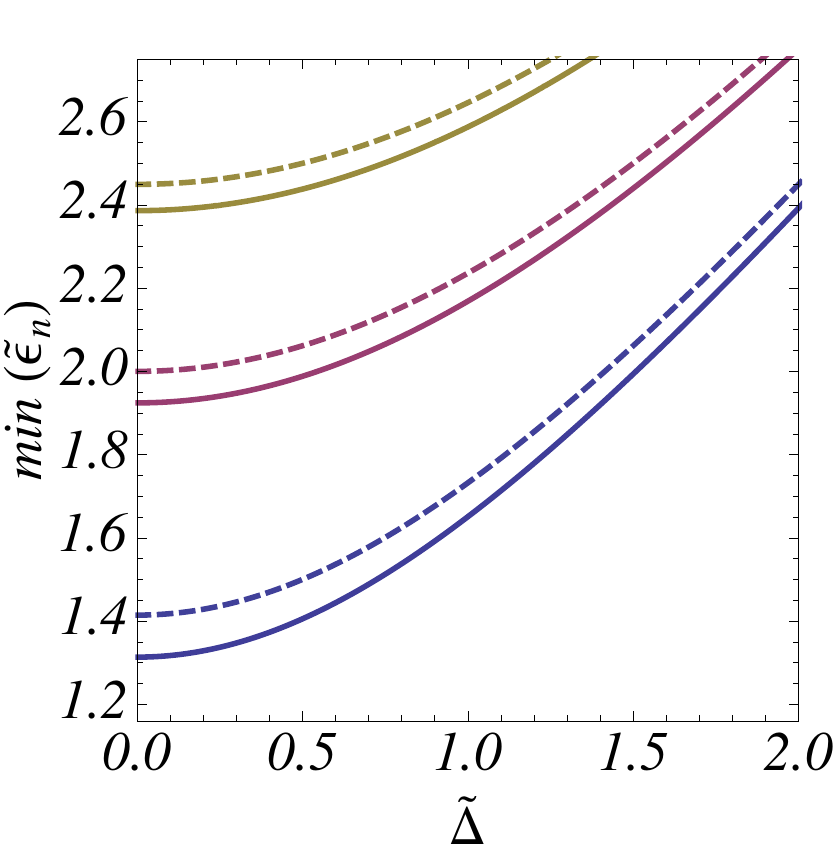}\,
\includegraphics[width=0.26\textwidth]{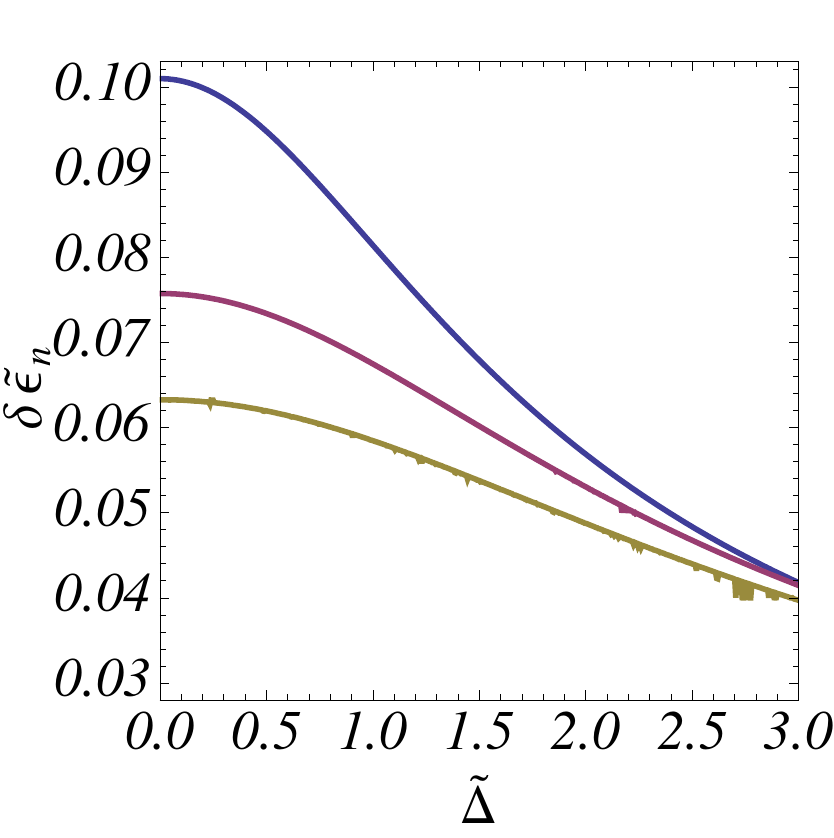}\,
\includegraphics[width=0.245\textwidth]{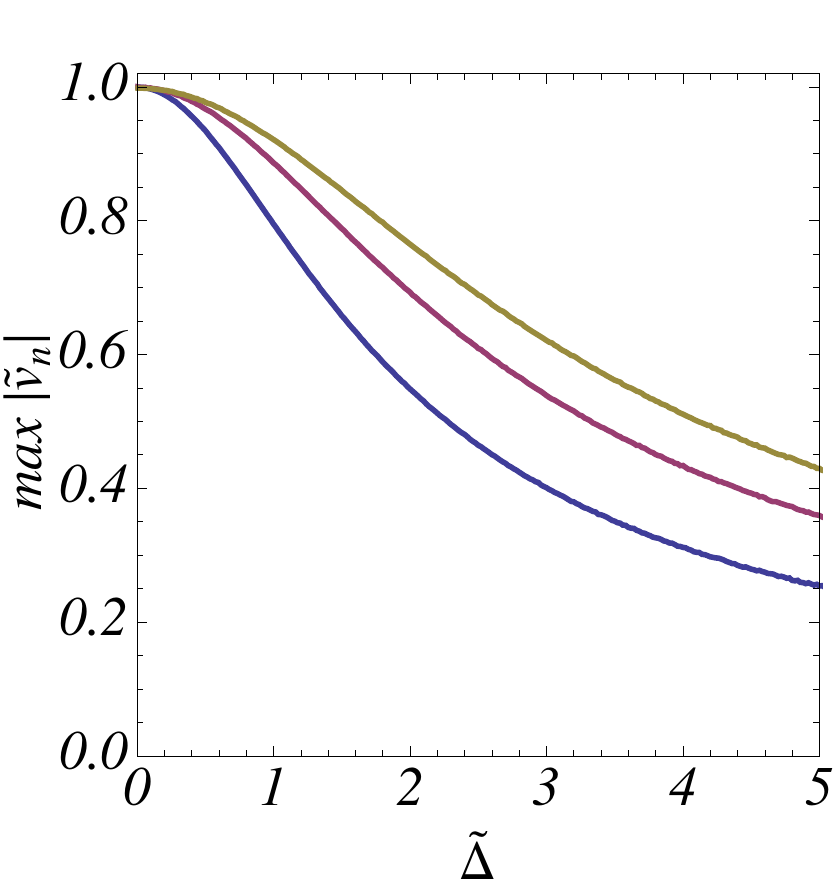}
\caption{(Left) Minima of the first three energy levels $\tilde \epsilon_n(k_y)$, with $n=1$ (blue solid line), $n=2$ (red solid line), $n=3$ (yellow solid line), associated to the edge modes with $k_y=p_n$, as functions of the mass term (solid lines). The dashed lines are the corresponding energy levels in the deep bulk embedded by a uniform magnetic field, described by the Landau levels $\tilde E_n=\sqrt{\tilde \Delta^2+2n}$. (Middle) Bandwidths of the first three levels, defined as the difference between the Landau levels, at the bulk, and the energy at the boundary, $\delta \tilde \epsilon_n= \tilde E_n-\min(\tilde \epsilon_n)$. (Right) Maximum velocities obtained at the threshold of the three levels, see Fig.~\ref{fig.en}, as functions of the mass term. In all the plots the blue lines correspond to $n=1$, the red lines to $n=2$, the yellow lines to $n=3$.}
\label{fig.D}
\end{figure}

 The dispersive energy levels are obtained by solving the matching condition Eq.~(\ref{mach_cond}). We verified that the bound states exists for $k_y<0$ and for 
 \be
 \tilde \epsilon^2\le \tilde k_y^2+\tilde \Delta^2
 \ee
 as shown in Fig.~\ref{fig.spectra}. For $k_y\rightarrow -\infty$ the energies $\tilde\epsilon_n$, solutions of Eq.~(\ref{mach_cond}), approach the Landau levels for relativistic massive particles 
 \be
 \tilde E_n=\pm \sqrt{\tilde \Delta^2+2n},
 \ee
 with $n$ positive integer numbers. In this limit the wavefunctions are written in terms of Hermite polynomials, as already mentioned. 
 For any $n$, $\tilde E_n$ is the maximum value of $\tilde \epsilon_n$. The minimum value of $\tilde \epsilon_n$ is located at the threshold,  
 $\tilde k_y=p_n\equiv -\sqrt{\tilde\epsilon_n^2-\tilde\Delta^2}$, solution of the equation
 \begin{equation}
 \label{pn_eq}
p_{n} \,D_{p_n^2/2-1}(\sqrt{2}\,p_n)-\sqrt{2}\,D_{p_n^2/2}(\sqrt{2}\,p_n)=0
\end{equation}
therefore $p_n$ does not depend on the mass term, as shown in Fig.~\ref{fig.en} (first two plots). 
For instance, numerically, we get $p_1\approx -1.31325$, $p_2\approx -1.92427$, $p_3\approx -2.38626$ and so on.
We have then
\be
\min\left[\tilde \epsilon_n(\tilde k_y)\right]=\tilde \epsilon_n(p_n)=\sqrt{\tilde \Delta^2+p_n^2}
\ee
see Fig.~\ref{fig.D} (first plot) where these quantities are reported as functions of the mass term $\tilde \Delta$. 
These bands are dispersive and the corresponding bandwidths can be easily calculated 
\be
\delta \tilde \epsilon_n= \tilde E_n-\tilde \epsilon_n(p_n)=\sqrt{\tilde \Delta^2+2n}-\sqrt{\tilde \Delta^2+p_n^2}
\ee
In Fig.~\ref{fig.D} (second plot) the bandwidths of the first three levels as functions of the mass $\tilde \Delta$ are reported. 
The wavefunctions associated to the dispersive part of the band $\tilde \epsilon_n$ are states localized at the edge of the magnetic field. These  edge states provide one-dimensional channels freely propagating along the magnetic boundary. Indeed, we can define the following velocities
\be
\tilde v_n(\tilde k_y)=\frac{d \tilde\epsilon_n}{d\tilde k_y}
\ee
and observe that their maximum absolute values are reached right at the threshold
\be
\max \big|\tilde v_n(\tilde k_y)\big|=|\tilde v_n(p_n)|
\ee
while for modes with $|k_y|>|p_n|$, $|\tilde v_n(\tilde k_y)|$ are smaller, see Fig.~\ref{fig.en} (last plot) for $n=1, 2$. In particular, for $\tilde \Delta=0$ we have $|\tilde v_n(p_n)|=1$, while increasing the mass term the modulus of the velocity decreases. 
Surprisingly we find that the ratios between the bandwidths and the maximum velocities, although both functions of the mass term, are universal quantities, namely, in their turn, the ratios do not depend on the mass, but are equal to the bandwidths in the massless case, at $\tilde\Delta=0$. Actually we observe numerically, at least for the first three levels reported in Fig.~\ref{fig.D}, that 
\be
\label{rescale}
\frac{\delta \tilde \epsilon_n[\tilde \Delta]}{\delta \tilde \epsilon_n[0]}=|\tilde v_n(p_n)|[\tilde \Delta]
\ee
We checked that the curves in the second and third plots of Fig.~\ref{fig.D} perfectly overlaps after rescaling according to Eq.~(\ref{rescale}).
This observation allows us to explicitly write the highest velocities as analytical functions of the mass parameter $\tilde \Delta$
\be
|\tilde v_n(p_n)|=\frac{1}{\sqrt{2n}-|p_n|}\left(\sqrt{\tilde \Delta^2+2n}-\sqrt{\tilde \Delta^2+p_n^2}\right),
\ee
with $p_n$ solution of Eq.~(\ref{pn_eq}). We finally notice that all curves representing $\tilde v_n(\tilde k_y)$ in log-scale reported in the last plot of Fig.~\ref{fig.D} collapse into a single curve after a rescaling, 
$\tilde v_n(\tilde k_y)[\tilde\Delta]=\tilde v_n(\tilde k_y)[0] |\tilde v_n(p_n)|[\tilde \Delta]$. Calling, for each band, $\tilde v_n^{o}(\tilde k_y)\equiv\tilde v_n(\tilde k_y)[\tilde\Delta=0]$ the velocity in the massless case, we have the following simple scaling law for the velocities in the massive ones
\be
\tilde v_n(\tilde k_y)=\frac{\tilde v_n^{o}(\tilde k_y)}{\sqrt{2n}-|p_n|}\left(\sqrt{\tilde \Delta^2+2n}-\sqrt{\tilde \Delta^2+p_n^2}\right).
\ee
\vspace{-0.3cm}
 \section{Conclusion}\label{section.four}

In this paper we derived the bound state spectrum for massive Dirac fermions in graphene subjected to a perpendicular magnetic field with a step function profile.  We showed that the energy levels approaches the relativistic Landau levels while the dispersive parts of the bands exhibit some universal behaviors. We find that the mass term modifies the bulk spectrum while reducing the number and the speed of the traveling modes at the border of the magnetic region, however the threshold of each bound states does not depends on the mass term and the ratio between the maximum propagating velocities and the bandwidths is also a mass-independent quantity. 
In conclusion, we show how, by tuning the mass term, one can control the speed of the edge modes traveling along the boundary of the magnetic region, paving the way for novel tunable graphene-based mesoscopic devices.

\vspace{-0.25cm}

\acknowledgements
We would like to thank  Reza Asgari, Alessandro De Martino, Ahmed Jellal for useful discussions.

\vspace{-0.3cm}


\begin{thebibliography}{99}

\bibitem{h1} A. H. Castro Neto, F. Guinea, N. M. R. Peres, K. S. Novoselov, A. K. Geim, Rev. Mod. Phys. 81, 109 (2009).
\bibitem{m1} G. W. Semenoff, Phys. Rev. Lett. 53 (1984) 2449.
\bibitem{m2} F. D. M. Haldane, Phys. Rev. Lett. 61, 2015 (1988).
\bibitem{m3}C.L. Kane, E.J. Mele, Phys. Rev. Lett. 95, 226801 (2005).
\bibitem{h2} N. M. R. Peres, A. H. Castro Neto, and F. Guinea, Phys. Rev. B 73,  241403(R) (2006).
\bibitem{h3} N. M. R. Peres and E. V. Castro, J. Phys.: Condens. Matter 19,  406231 (2007).
\bibitem{h4} M. Farjam and H. Rafii-Tabar, Phys. Rev. B 79,  045417 (2009).
\bibitem{h5} S. Kuru, J. Negro, and L. M. Nieto, J. Phys. Condens. Matter 21, 455305 (2009).
\bibitem{h6} C. G. Beneventano and E. M. Santangelo, J. Phys. A: Math. Theor.   39,  7457 (2006).
\bibitem{h7} G. Giovannetti, P. A. Khomyakov, G. Brocks, P. J. Kelly, J. van den Brink, Phys. Rev. B 76,  073103 (2007).
\bibitem{h8} V. P. Gusynin and S. G. Sharapov, Phys. Rev. Lett. 95,  146801 (2005).
\bibitem{h9} M. O. Goerbig, Rev. Mod. Phys. 83, 1193 (2011).
\bibitem{h10} B. Midya and D. J. Fernndez, J. Phys. A: Math. Theor. 47, 28, 285302 (2014).
\bibitem{h11} J. M. Pereira Jr, V. Mlinar, F. M. Peeters, and P. Vasilopoulos, Phys. Rev. B 74,  045424 (2006).
\bibitem{h12} M. R. Setare and D. Jahani, Physica B 405,  1433 (2010).
\bibitem{h14} A. De Martino, L. Dell'Anna, and R. Egger, Solid State Commun. 144,  547 (2007).
\bibitem{h14b} T. K. Ghosh, A. De Martino, W. H\"ausler, L. Dell'Anna, and R. Egger, Phys. Rev. B.77, 081404(R) (2008).
\bibitem{h13} M. R. Masir, P. Vasilopoulos, A. Matulis, and F. M. Peeters, Phys. Rev. B.77, 235443 (2008).
\bibitem{h15} A. Kormanyos, P. Rakyta, L. Oroszlany, and J. Cserti, Phys. Rev. B 78, 045430 (2008).
\bibitem{h16} A. De Martino, L. Dell'Anna, and R. Egger, Phys. Rev. Lett. 98,  066802 (2007).
\bibitem{h17} M. R. Masir, P. Vasilopoulos, and F. M. Peeters, New J. Phys. 11,  095009 (2009).
\bibitem{h18} N. Myoung, G. Ihm, S. J. Lee, Physica E: Low-dimensional Systems and Nanostructures, 42, 10, 2808 (2010).
\bibitem{h19} E. B. Choubabi, M. E. Bouziani, and A. Jellal, Int. J. Geom. Meth. Mod. Phys. 7,  909 (2010).
\bibitem{h20} E. Milpas, M. Torres, and G. Murguia, J. Phys.: Condens. Matter 23,  245304 (2011).
\bibitem{h21} L. Dell'Anna and A. De Martino, Phys. Rev. B 79,  045420 (2009).
\bibitem{h22} S. Park and H. S. Sim, Phys. Rev. B 77, 075433 (2008).
\bibitem{h23} M. R. Masir, P. Vasilopoulos, and F. M. Peeters, Appl. Phys. Lett. 93,  242103 (2008).
\bibitem{h24} S. Ghosh and M. Sharma, J.  Phys.: Condens. Matter 21,  292204 (2009).
\bibitem{h25} N. Myoung and G. Ihm, Physica E 42, 70 (2009).
\bibitem{h26} N. Agrawal, S. Chosh, and M. Sharma, I. J. Mod. Phys. B 27,  1341003 (2013).
\bibitem{h27} M. Esmailpour, Physica B 534, 150 (2018).

\end{thebibliography}
\end{document}